\documentclass[aps,pra,twoside,twocolumn,showpacs,floatfix,pdflatex]{revtex4-1}

\usepackage{graphicx}
\usepackage{color}
\usepackage{latexsym}

\begin{document}

\title{Increasing efficiency of a linear-optical quantum gate using an
electronic feed forward}

\author{Martina Mikov\'{a}}
\author{Helena Fikerov\'{a}}
\author{Ivo Straka}
\author{Michal Mi{\v{c}}uda}
\author{Miroslav Je\v{z}ek}
\author{Miloslav Du{\v{s}}ek}

\affiliation{Department of Optics, Faculty of Science, Palacky University,
             17.~listopadu 12, 771\,46 Olomouc, Czech Republic}

\begin{abstract}
We have successfully used a fast electronic feed forward to increase the success
probability of a linear optical implementation of a programmable phase gate from
25\,\% to its theoretical limit of 50\,\%. The feed forward applies a
conditional unitary operation which changes the incorrect output states of the data
qubit to the correct ones. The gate itself rotates an arbitrary quantum state of
the data qubit around the $z$-axis of the Bloch sphere with the angle of rotation
being fully determined by the state of the program qubit. The gate
implementation is based on fiber optics components. Qubits are encoded into
spatial modes of single photons. The signal from the feed-forward detector is led
directly to a phase modulator using only a passive voltage divider. We have
verified the increase of the success probability and characterized the gate
operation by means of quantum process tomography. We have demonstrated that the
use of the feed forward does not affect either the process
fidelity or the output-state fidelities.
\end{abstract}


\maketitle

\section{Introduction}

Linear-optical architectures belong to the most prominent platforms for realizing protocols of quantum information processing \cite{kok07,kni01}. They are experimentally feasible and they work directly with photons without the necessity to transfer the quantum state of a photonic qubit into another quantum system like an ion etc. The latter feature is quite convenient because photons are good carriers of information for communication purposes. Linear-optical quantum gates achieve the non-linearity necessary for the interaction between qubits by means of the non-linearity of quantum measurement. Unfortunately, quantum measurement is not only non-linear but also probabilistic. Therefore linear-optical implementations of quantum gates are mostly probabilistic too---their operation sometimes fails. Partly, this is a fundamental limitation. But in many cases when data qubits appear in an improper state after the measurement on an ancillary system they can still be corrected by applying a proper unitary transformation which depends only on the measurement result. In these situations, implementation of the feed forward can increase the probability of success of the gate \cite{sci06,pre07}. In the present paper, we apply this approach to a linear-optical programmable quantum gate.

The most of conventional computers use fixed hardware and different tasks are performed using different software. This concept can be, in principle, applied to quantum computers as well: The executed unitary operation can be determined by some kind of a program. However, in 1997 Nielsen and Chuang \cite{nie97} showed that an $n$-qubit quantum register can perfectly encode at most $2^n$ distinct quantum operations. Although this bound rules out perfect universally-programmable quantum gates (even unitary transformations on only one qubit form a group with uncountably many elements), it is still possible to construct approximate or probabilistic programmable quantum gates and optimize their performance for a given size of the program register. Such gates can either operate deterministically, but with some noise added to the output state \cite{hil06}, or they can operate probabilistically, but error free \cite{hil02,vid02,hil04}. Combination of these regimes is also possible.

A probabilistic programmable phase gate was proposed by Vidal, Masanes, and Cirac \cite{vid02}. It carries out rotation of a single-qubit state along the $z$-axis of the Bloch sphere. The angle of rotation (or the phase shift) is programmed into a state of a single-qubit program register. It is worth noting that an exact specification of an angle of rotation would require infinitely many classical bits. But here the information is encoded into a single qubit only. The price to pay is that the success probability of such a gate is limited by 50\,\% \cite{note1}. The programmable phase gate was experimentally implemented for the first time in 2008 \cite{mic08}. However, the success probability of that linear-optical implementation reached only 25\,\%. In the present paper we will show how to increase the success probability of this scheme to its quantum mechanical limit of 50\,\% by means of electronic feed forward.

\section{Theory}

The programmable phase gate works with a data and program qubit. The program qubit is supposed to contain information about the phase shift $\phi$ encoded in the following way:
\begin{equation}
|\phi\rangle_{P}=\frac{1}{\sqrt{2}}(|0\rangle_P + e^{i\phi}|1\rangle_P).
\label{eq-prog_qubit}
\end{equation}
The gate performs a unitary evolution of the data qubit which depends on the state of the program qubit:
\begin{equation}
U(\phi) =|0\rangle_D \langle 0|+e^{i\phi}|1\rangle_D \langle 1|.
\label{eq-Uphi}
\end{equation}
Without loss of generality we can consider only pure input states of the data qubit:
\begin{equation}
|\psi_\mathrm{in}\rangle_{D} = \alpha|0\rangle_D+\beta|1\rangle_D.
\label{eq-dat_qubit_in}
\end{equation}
So the output state of the data qubit reads:
\begin{equation}
|\psi_\mathrm{out}\rangle_{D} = \alpha|0\rangle_D + e^{i\phi}\beta|1\rangle_D.
\label{eq-dat_qubit_out}
\end{equation}

\begin{figure}
  \begin{center}
  \resizebox{\hsize}{!}{\includegraphics*{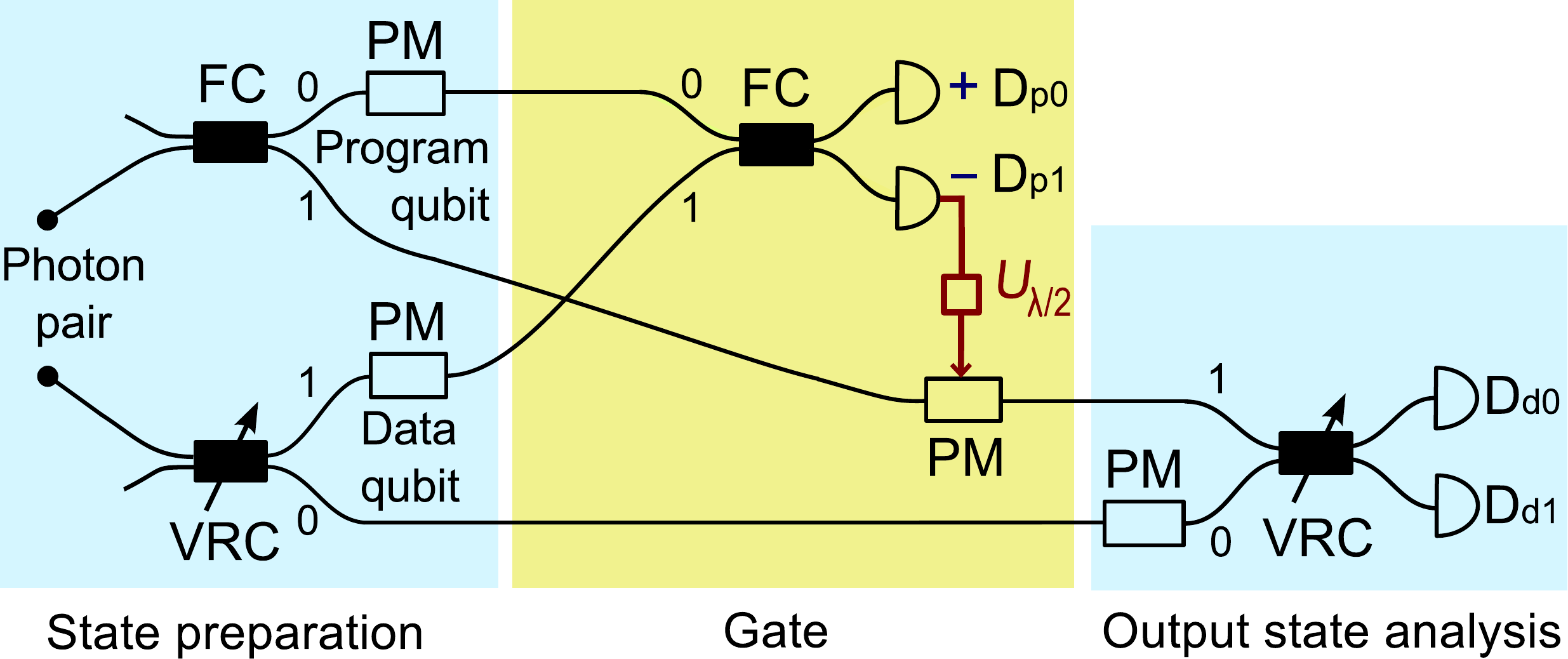}}
  \end{center}
  \caption{(Color online) Scheme of the experiment. FC -- fiber couplers,
           VRC -- variable ratio couplers, PM -- phase modulators, D -- detectors.}
  \label{fig-scheme}
\end{figure}

Experimentally the programmable phase gate can be implemented by an optical setup shown in Fig.~\ref{fig-scheme}. Each qubit is represented by a single photon which may propagate in two optical fibers. The basis states $|0\rangle$ and $|1\rangle$ correspond to the presence
of the photon in the first or second fiber, respectively. When restricted only to the cases where a single photon emerges in each output port, the conditional two-photon output state reads
(the normalization reflects the fact that the probability of this situation is $1/2$):
\begin{eqnarray*}
&&\hspace{-5mm}
\frac{1}{\sqrt{2}}(\alpha|0\rangle_D \otimes |0\rangle_P+\beta e^{i\phi} |1\rangle_D \otimes |1\rangle_P)\\
&=& \frac{1}{2} \big[ (\alpha|0\rangle_D + \beta e^{i\phi}|1\rangle_D) \!\otimes\! |+\rangle_P\\
&+& (\alpha|0\rangle_D - \beta e^{i\phi}|1\rangle_D) \!\otimes\! |-\rangle_P \big],
\end{eqnarray*}
where $|\pm\rangle_P=\frac{1}{\sqrt{2}}(|0\rangle_P \pm |1\rangle_P)$. If we make a measurement on the program qubit in the basis $\{| \pm \rangle_P \}$ then also the output state of the data qubit collapses into one of the two following states according to the result of the measurement: $|\psi_{\mathrm{out}}\rangle_D=\alpha|0\rangle_D \pm \beta e^{i\phi}|1\rangle_D$.
If the measurement outcome is $|+\rangle_P$ then the unitary transformation $U(\phi)$ has been applied to the data qubit. If the outcome is $|-\rangle_P$ then the state acquires an extra $\pi$ phase shift, i.e., $U(\phi+\pi)$ has been executed. This is compensated by a fast electro-optical modulator which applies a corrective phase shift $-\pi$ (in practice we apply phase shift $\pi$ what is equivalent).

\section{Experiment}

The scheme of the setup is shown in Fig.~\ref{fig-scheme}. Pairs of photons are created by type-II collinear frequency-degenerate spontaneous parametric down conversion (SPDC) in a two-millimeter long BBO crystal pumped by a diode laser (Coherent Cube) at 405\,nm. The photons are separated by a polarizing beam splitter and coupled into single-mode fibers. Using fiber polarization controllers the same polarizations are set on the both photons. By means of fiber couplers and electro-optical phase modulators the required input states of the program and data qubits are prepared. To prepare state (\ref{eq-prog_qubit}) of the program qubit a fiber coupler (FC) with fixed splitting ratio 50:50 is used. An arbitrary state of the data qubit (\ref{eq-dat_qubit_in}) is prepared using electronically controlled variable ratio coupler (VRC). All employed phase modulators (EO Space) are based on the linear electro-optic effect in lithium niobate. Their half-wave voltages are about 1.5\,V. These phase modulators (PM) exhibit relatively high dispersion. Therefore one PM is placed in each interferometer arm in order to compensate dispersion effects. Because the overall phase of a quantum state is irrelevant it is equivalent to apply either a phase shift $\varphi$ to $|1\rangle$ or $-\varphi$ to $|0\rangle$.

The gate itself consists of the exchange of two rails of input qubits and of the measurement on the data qubit (see Fig.~\ref{fig-scheme}). The measurement in basis $\{| \pm \rangle \}$ is accomplished by a fiber coupler with fixed splitting ratio 50:50 and two single photon detectors. In this experiment we use single photon counting modules (Perkin-Elmer) based on actively quenched silicon avalanche photodiodes. Detectors D$_{p0}$, D$_{d0}$, and D$_{d1}$ belongs to a quad module SPCM-AQ4C (total efficiencies 50--60\,\%, dark counts 370--440\,s$^{-1}$, response time 33--40\,ns). As detector D$_{p1}$, serving for the feed forward, a single module SPCM AQR-14FC is used because of its faster response (total efficiency about 50\,\%, dark counts 180\,s$^{-1}$, response time 17\,ns). The output of the detector is a 31\,ns long TTL (5\,V) pulse.

To implement the feed forward the signal from detector D$_{p1}$ is led to a passive voltage divider in order to adapt the 5\,V voltage level to about 1.5\,V (necessary for the phase shift of $\pi$) and then it is led directly to the phase modulator. The coaxial jumpers are as short as possible. The total delay including the time response of the detector is 20\,ns. To compensate this delay, photon wave-packets representing data qubits are retarded by fiber delay lines (one coil of fiber of the length circa 8\,m in each interferometer arm). Timing of the feed-forward pulse and the photon arrival was precisely tuned. Coherence time of photons created by our SPDC source is only several hundreds of femtoseconds.

The right-most block in Fig.~\ref{fig-scheme} enables us to measure the data qubit at the output of the gate in an arbitrary basis. These measurements are necessary to evaluate performance of the gate.

The whole experimental setup is formed by two Mach-Zehnder interferometers (MZI). The length of the arms of the shorter MZI is about 10.5\,m (the upper interferometer in Fig.~\ref{fig-scheme}). The length of the arms of the longer one is about 21.5\,m (the lower interferometer one in Fig.~\ref{fig-scheme}). To balance the arm lengths we use motorized air gaps with adjustable lengths. Inside the air gaps, polarizers and wave plates are also mounted. They serve for accurate setting of polarizations of the photons (to obtain high visibilities the polarizations in the both arms of each MZI must be the same).

To reduce the effect of the phase drift caused by fluctuations of temperature and temperature gradients we apply both passive and active stabilization. The experimental setup is covered by a shield minimizing air flux around the components and the both delay fiber loops are winded on an aluminium cylinder which is thermally isolated. Besides, after each three seconds of measurement an active stabilization is performed. It measures intensities for phase shifts 0 and $\pi/2$ and if necessary it calculates phase compensation and applies corresponding additional corrective voltage to the phase modulator. This results in the precision of the phase setting during the measurement period better than $\pi/200$. For the stabilization purposes we use a laser diode at 810\,nm. To ensure the same spectral range, both the laser beam and SPDC generated photons pass through the same band-pass interference filter (spectral FWHM 2\,nm, Andover). During the active stabilization the source is automatically switched from SPDC to a laser diode.

\section{Results}

Any quantum operation can be fully described by a completely
positive (CP) map. According to the Jamiolkowski-Choi isomorphism
any CP map can be represented by a positive-semidefinite operator
$\chi$ on the tensor product of input and output Hilbert spaces
$\mathcal{H}_{\mathrm{in}}$ and $\mathcal{H}_{\mathrm{out}}$
\cite{jam72,cho75}. The input state $\rho_{\mathrm{in}}$ transforms
according to
$$
\rho_{\mathrm{out}}= \mathrm{Tr}_{\mathrm{in}}[\chi
(\rho_{\mathrm{in}}^T\otimes I_{\mathrm{out}})].
$$
Combinations of different input states with measurements on the output quantum
system represent effective measurements performed on
$\mathcal{H}_{\mathrm{in}}\otimes\mathcal{H}_{\mathrm{out}}$.
A proper selection of the input states and final measurements
allows us to reconstruct matrix $\chi$ from measured data
using maximum likelihood (ML) estimation technique \cite{jez03,par04}.

\begin{figure}
  \begin{center}
   Reconstructed: \hspace{24mm} Ideal: \hspace*{11mm} \\[1mm]
   \resizebox{0.49\hsize}{!}{\includegraphics*{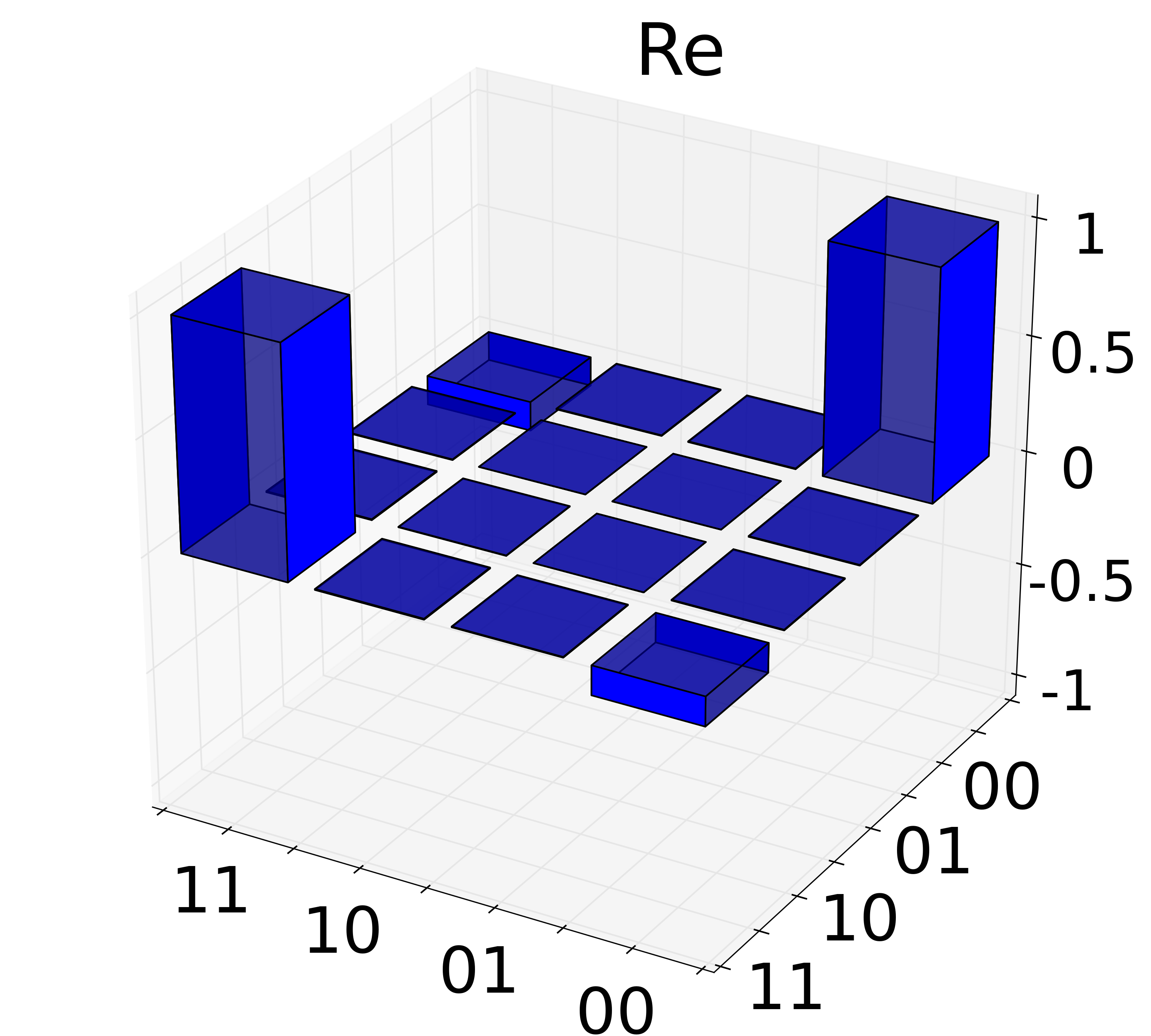}}
   \resizebox{0.49\hsize}{!}{\includegraphics*{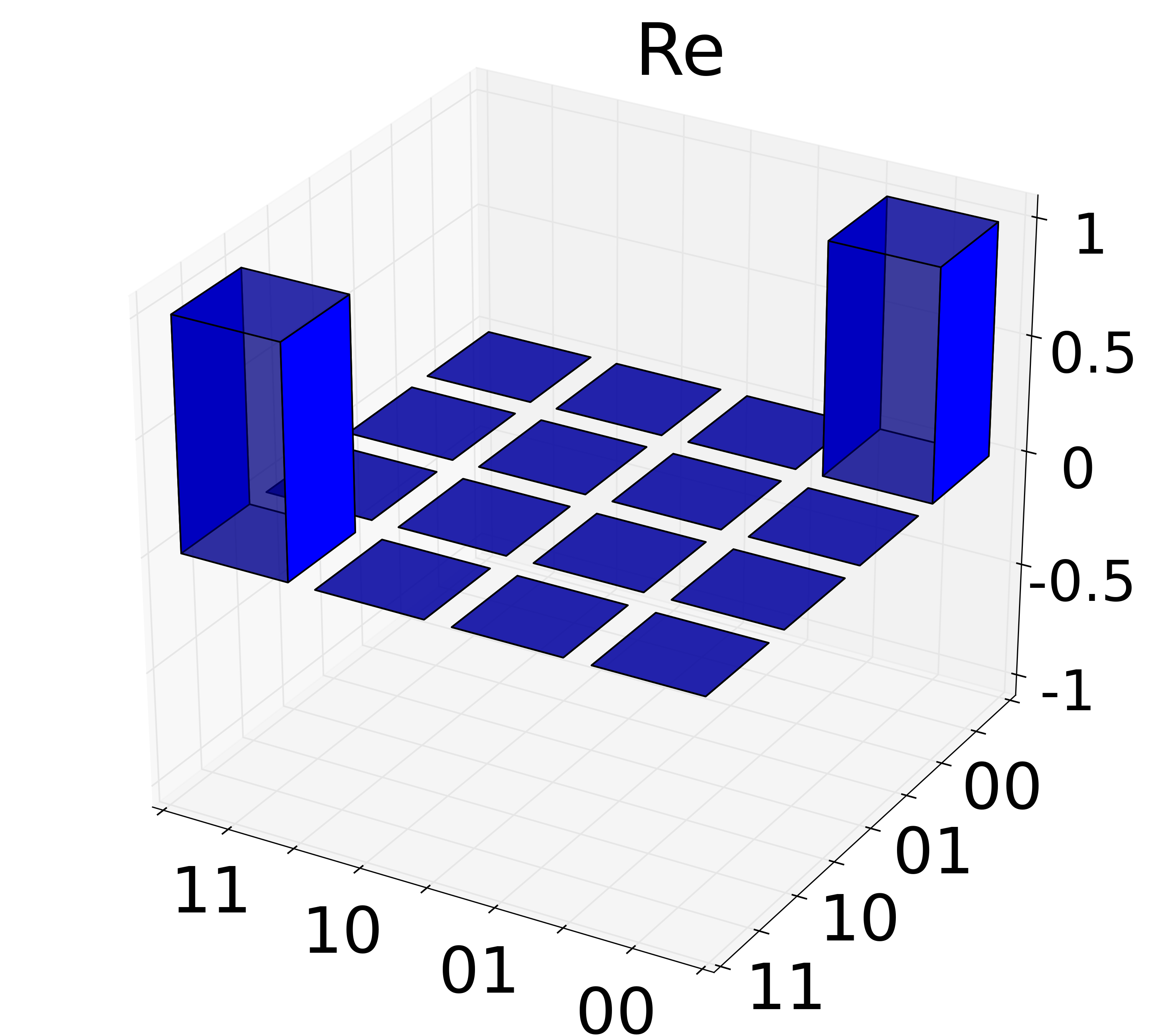}}\\[5mm]
   \resizebox{0.49\hsize}{!}{\includegraphics*{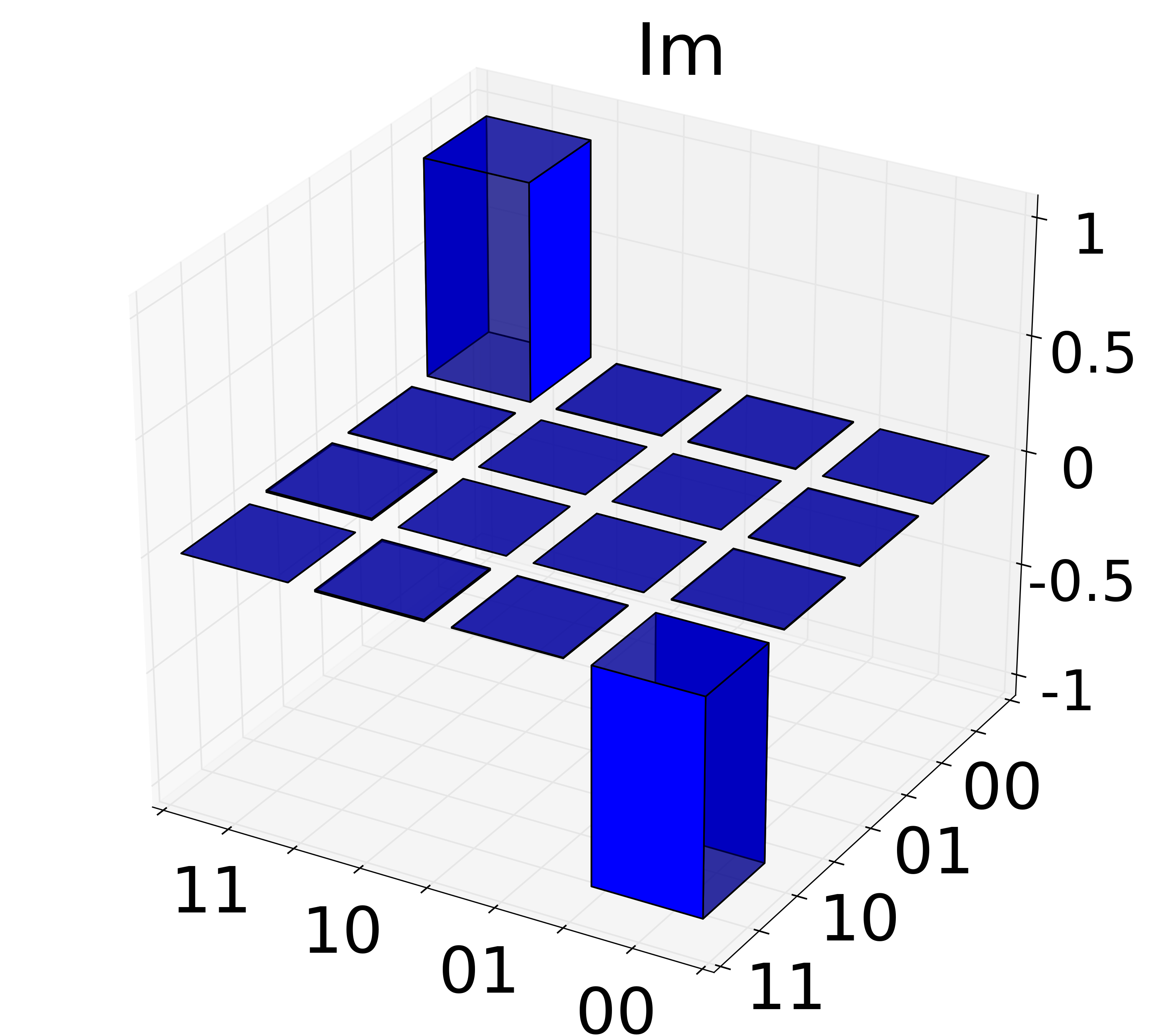}}
   \resizebox{0.49\hsize}{!}{\includegraphics*{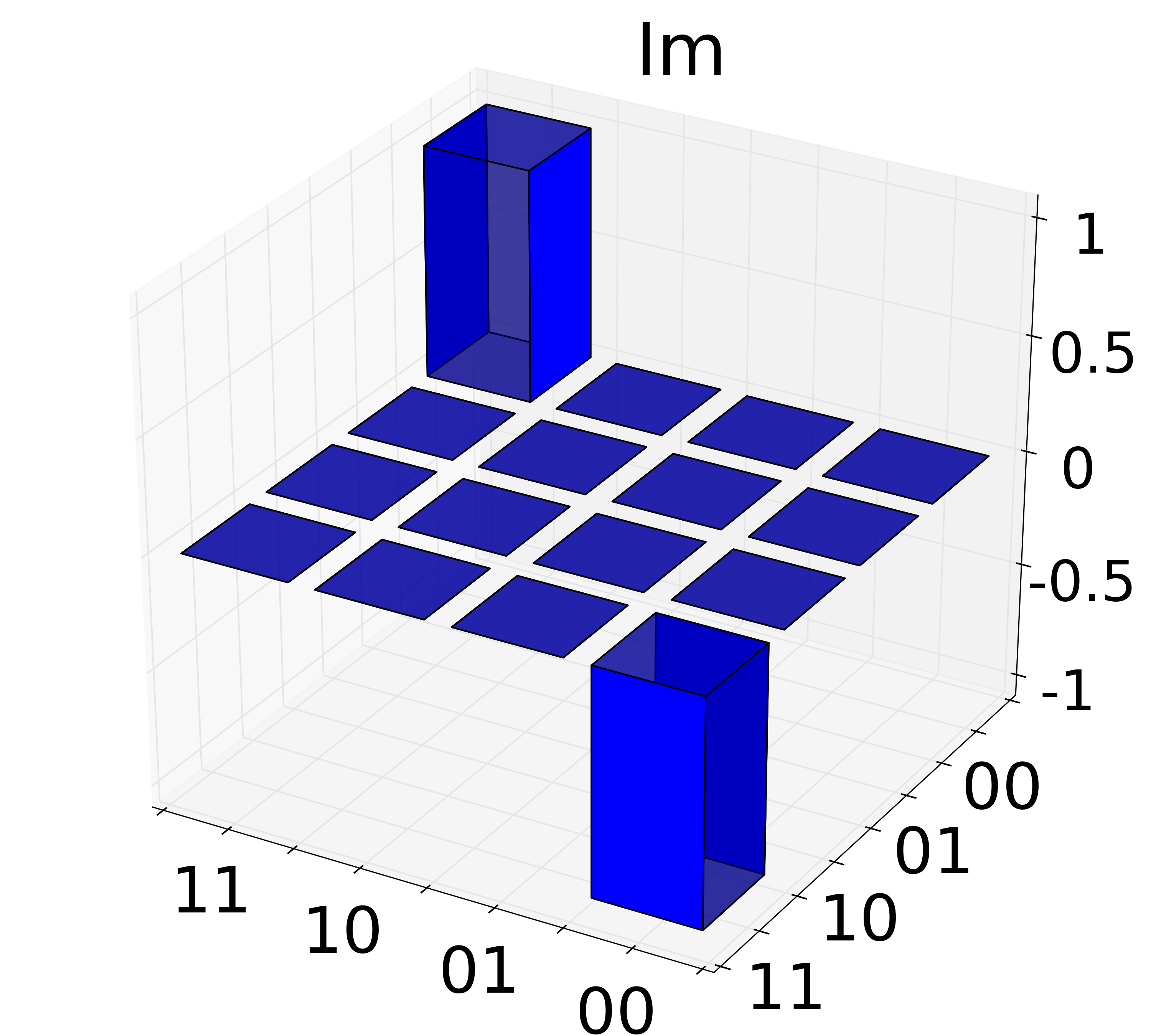}}
  \end{center}
  \caption{(Color online) Choi matrix for the gate with the feed forward when
           $\phi = \pi/2$ is encoded into the program qubit. The left top
           panel shows the real part of the reconstructed process matrix while
           the left bottom one displays its imaginary part.
           The two right panels show the real and imaginary part
           of the ideal matrix.}
  \label{fig-choi-halfpi}
\end{figure}

\begin{figure}
  \begin{center}
   Reconstructed: \hspace{24mm} Ideal: \hspace*{11mm} \\[1mm]
   \resizebox{0.49\hsize}{!}{\includegraphics*{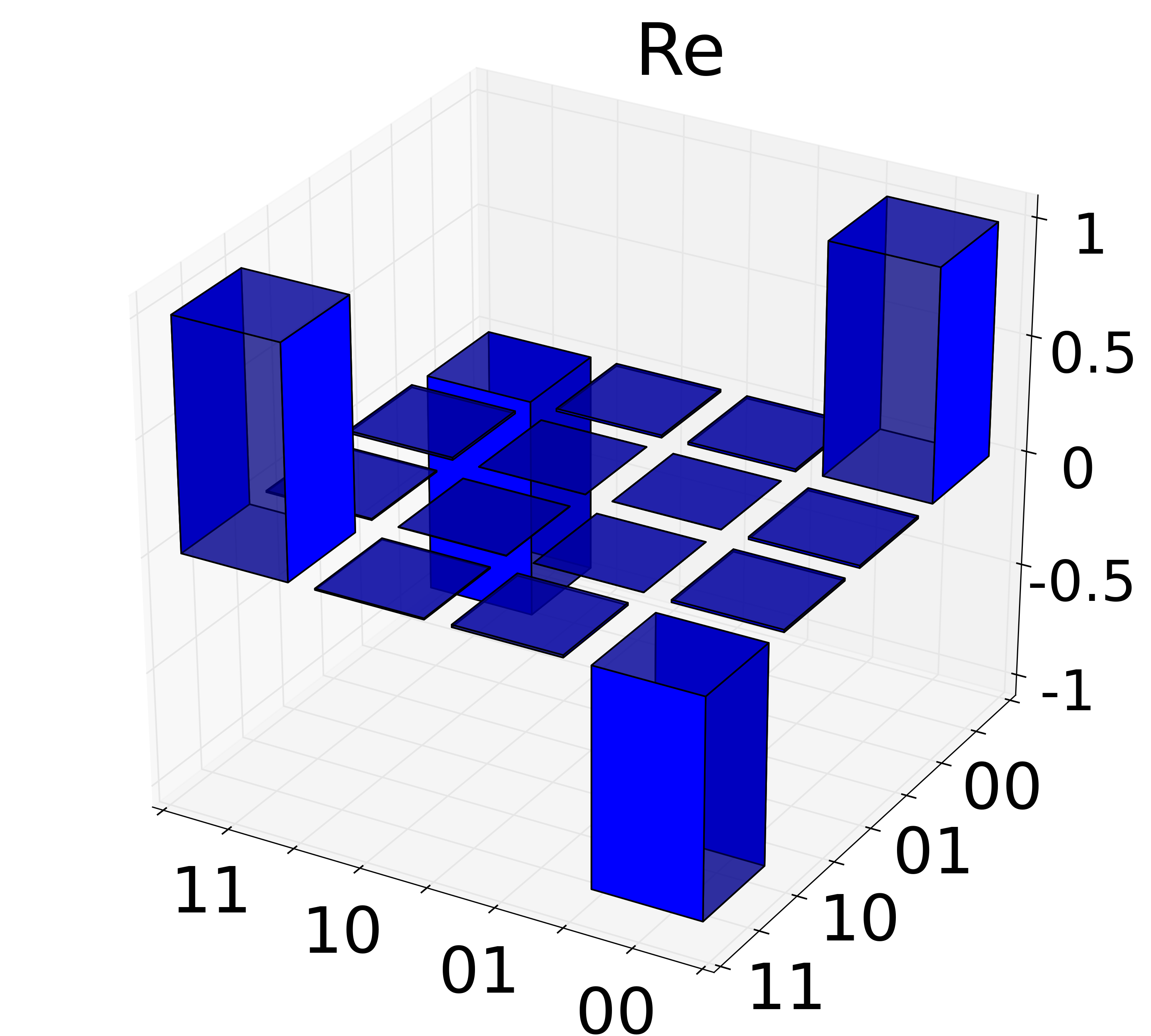}}
   \resizebox{0.49\hsize}{!}{\includegraphics*{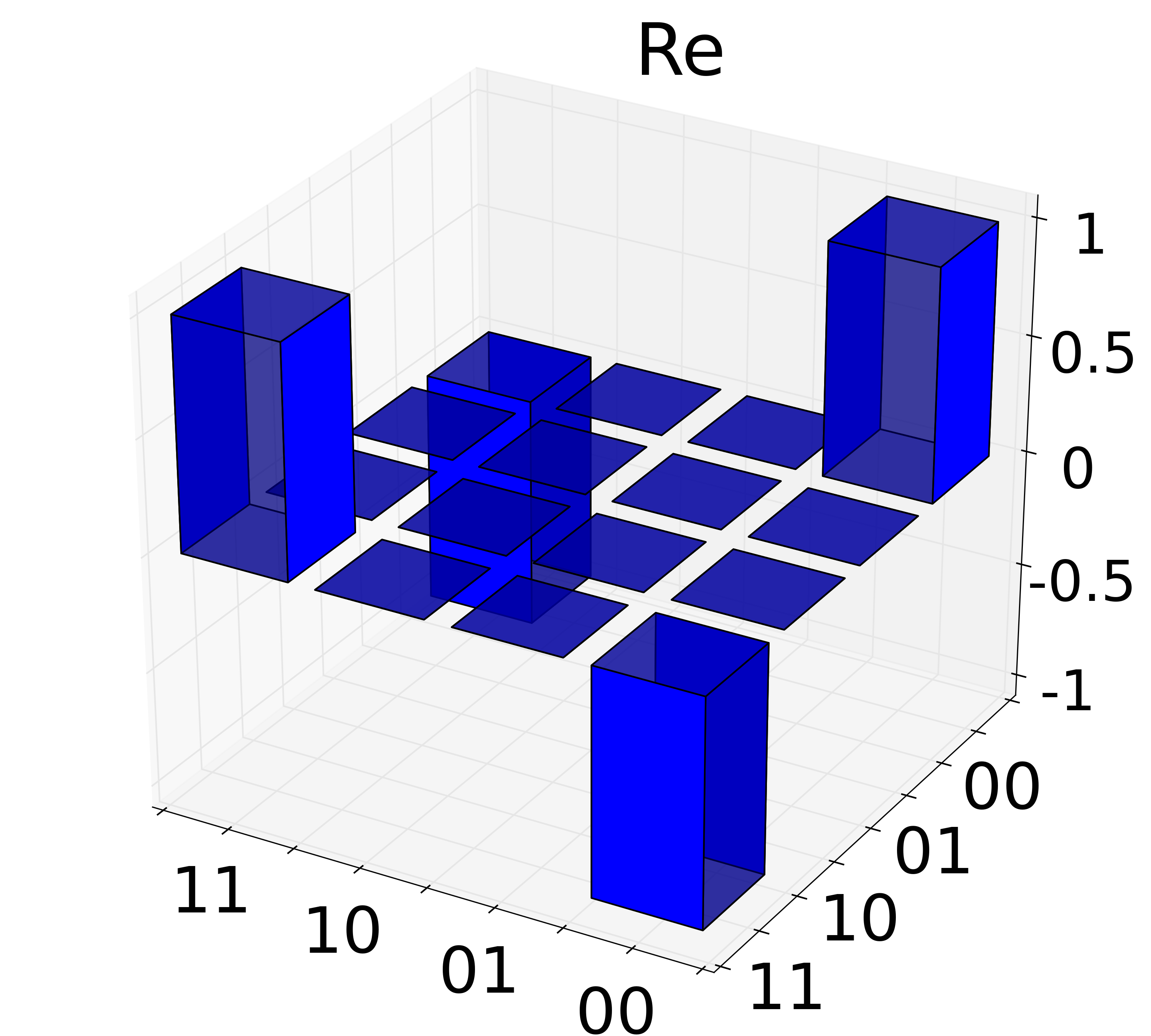}}\\[5mm]
   \resizebox{0.49\hsize}{!}{\includegraphics*{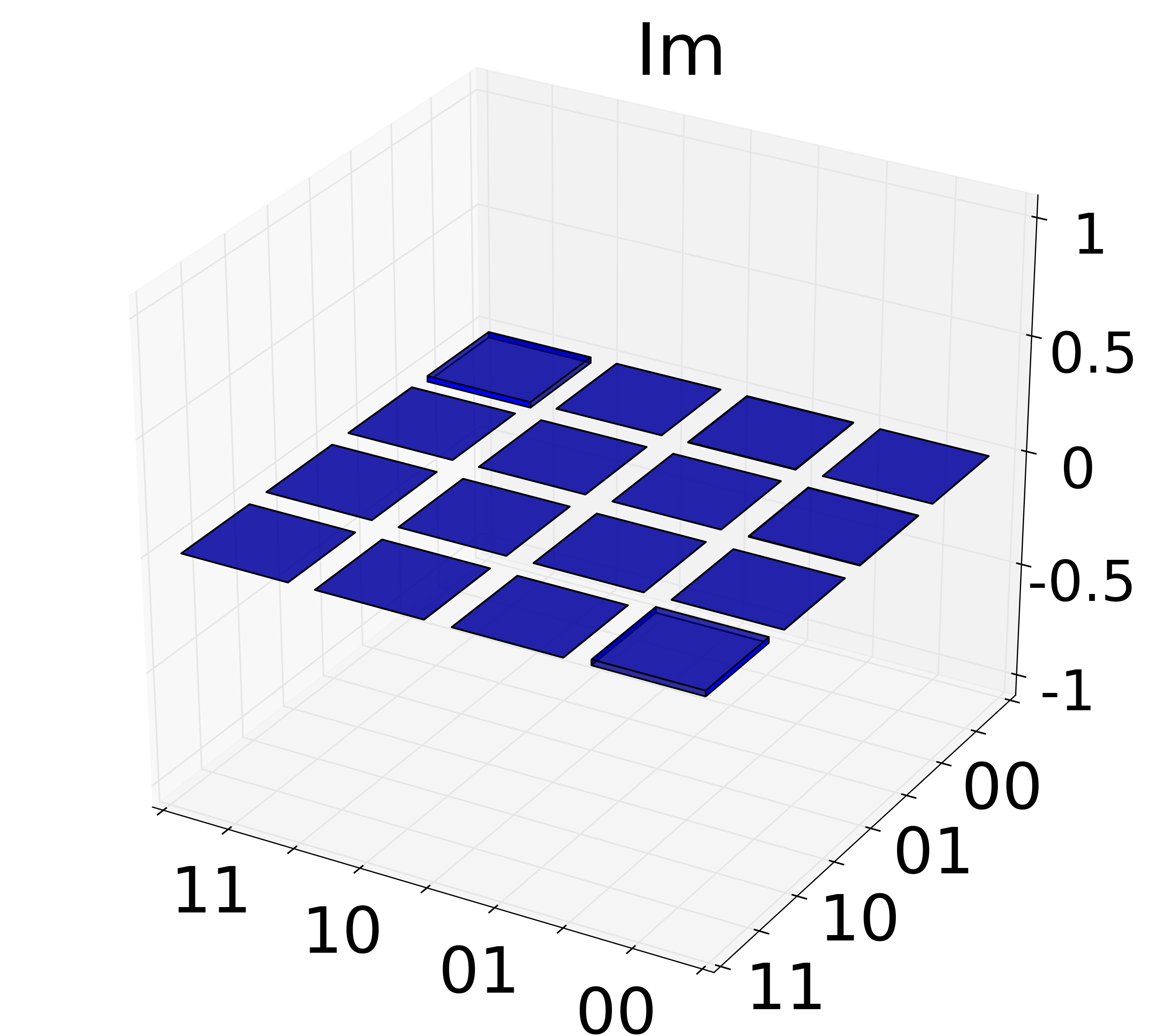}}
   \resizebox{0.49\hsize}{!}{\includegraphics*{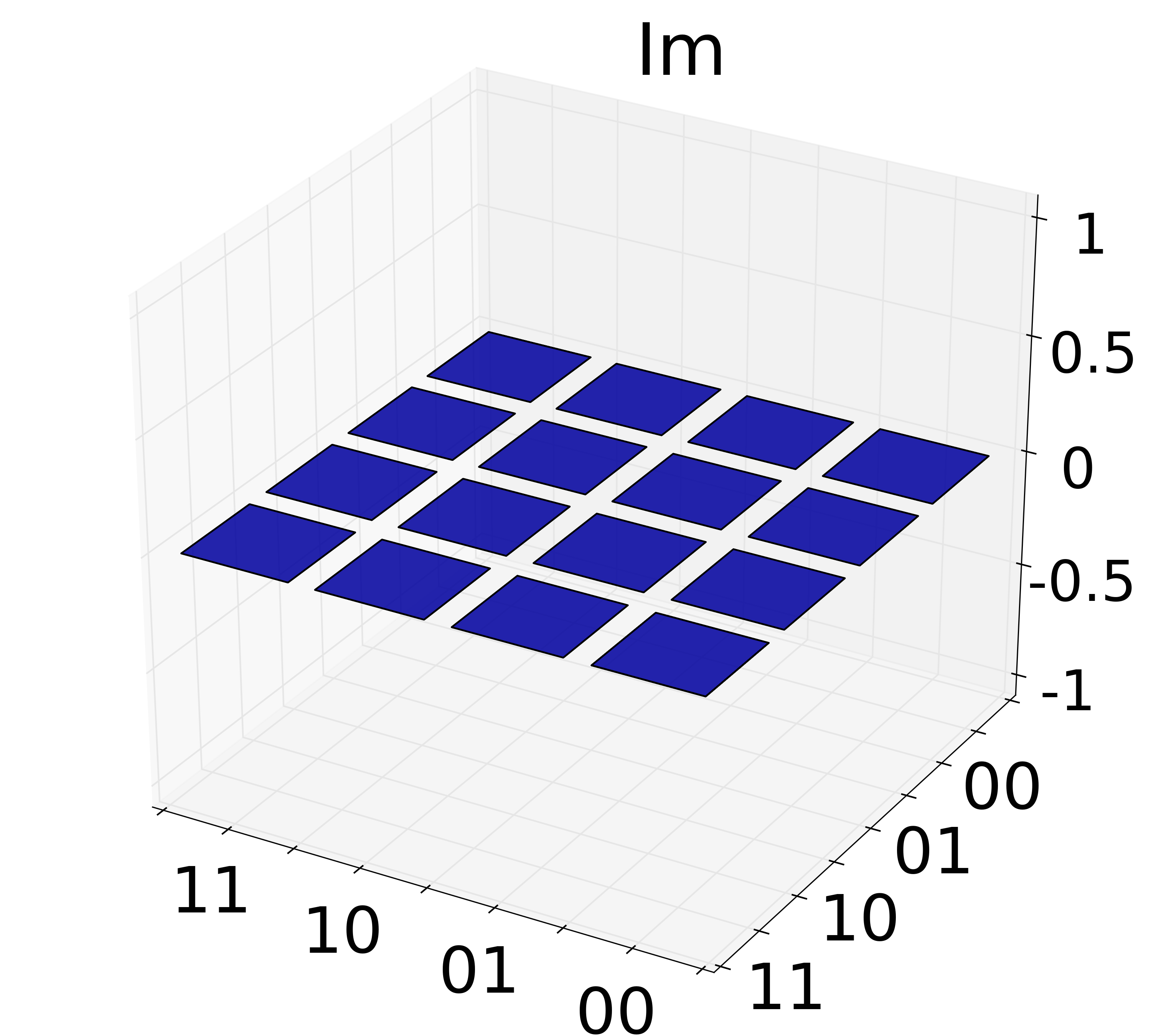}}
  \end{center}
  \caption{(Color online) Choi matrix for the gate with the feed forward when
           $\phi = \pi$ is encoded into the program qubit. The left top
           panel shows the real part of the reconstructed process matrix while
           the left bottom one displays its imaginary part.
           The two right panels show the real and imaginary part
           of the ideal matrix.}
  \label{fig-choi-pi}
\end{figure}

For each phase shift, i.e, for a fixed state of the program qubit, we used six different input states of the data qubit, namely
$|0\rangle, |1\rangle, |\pm\rangle = (|0\rangle \pm |1\rangle)/\sqrt{2},$ and
$|\pm i \rangle = (|0\rangle \pm i |1\rangle)/\sqrt{2}$.
For each of these input states the state of the data qubit at the output of the gate was measured
in three different measurement basis, $\{|0\rangle, |1\rangle \},
\{ |\pm \rangle \},$ and $\{ |\pm i \rangle \}$.
Each time we simultaneously measured two-photon coincidence counts between detectors
D$_{p0}$ \& D$_{d0}$, D$_{p0}$ \& D$_{d1}$, D$_{p1}$ \& D$_{d0}$, D$_{p1}$ \& D$_{d1}$
in 12 three-second intervals. The unequal detector efficiencies were compensated by proper rescaling of the measured coincidence counts.
From these data we have reconstructed Choi matrices describing the functioning of the gate for several different phase shifts. In Figs.~\ref{fig-choi-halfpi} and \ref{fig-choi-pi} there are  examples of the Choi matrices of the gate for $\phi=\pi/2$ and $\phi=\pi$, respectively.

To quantify the quality of gate operation we have calculated the process fidelity. If $\chi_{\mathrm{id}}$ is a one-dimensional projector then the common definition of process fidelity is
$$
F_\chi=\mathrm{Tr}[\chi \chi_{\mathrm{id}}] /
(\mathrm{Tr}[\chi]\mathrm{Tr}[\chi_{\mathrm{id}}]).
$$
Here $\chi_{\mathrm{id}}$ represents the ideal transformation
corresponding to our gate. In particular,
$$
\chi_{\mathrm{id}} = \sum_{i,j=0,1} |i \rangle\langle j|
\otimes U | i \rangle\langle j | U^\dag,
$$
where $U$ stays for the unitary operation (\ref{eq-Uphi}) applied by the gate.

We have also reconstructed density matrices of the output
states of the data qubit corresponding to all input states and calculated fidelities and purities of the output states. The fidelity of output state $\rho_{\mathrm{out}}$ is defined as
$F=\langle \psi_{\mathrm{out}}| \rho_{\mathrm{out}}
|\psi_{\mathrm{out}} \rangle,$
where $|\psi_{\mathrm{out}} \rangle = U |\psi_{\mathrm{in}} \rangle$
with $|\psi_{\mathrm{in}} \rangle$ being the (pure) input state.
The purity of the output state is defined as
$\mathcal{P}=\mathrm{Tr}[\rho_{\mathrm{out}}^2]$. If the input
state is pure the output state is expected to be pure as well.

Table~\ref{tab-withFF} shows process fidelities for seven different phase shifts. It also shows the average and minimal values of output state fidelities and average and minimal purities of output states. Fidelities and purities are averaged over six output states corresponding to six input states described above. Also the minimum values are related to these sets of states.

\begin{table}
 \begin{ruledtabular}
  \begin{tabular}{cccccc}
  $\phi$ & $F_{\chi}$ & $F_{\mathrm{av}}$ & $F_{\mathrm{min}}$ &
  $\mathcal{P}_{\mathrm{av}}$ & $\mathcal{P}_{\mathrm{min}}$
  \\ \hline
  0         & 0.976 & 0.985 & 0.970 & 0.974 & 0.947 \\
  $\pi/6$   & 0.977 & 0.986 & 0.972 & 0.975 & 0.951 \\
  $\pi/3$   & 0.977 & 0.985 & 0.970 & 0.975 & 0.943 \\
  $\pi/2$   & 0.974 & 0.983 & 0.973 & 0.975 & 0.953 \\
  $2\pi/3$  & 0.978 & 0.987 & 0.962 & 0.988 & 0.961 \\
  $5\pi/6$  & 0.972 & 0.981 & 0.953 & 0.974 & 0.944 \\
  $\pi$     & 0.980 & 0.987 & 0.975 & 0.977 & 0.961 \\
  \end{tabular}
 \end{ruledtabular}
 \caption{Process fidelities ($F_{\chi}$), average ($F_{\mathrm{av}}$)
          and minimal ($F_{\mathrm{min}}$) output-state fidelities,
          average ($\mathcal{P}_{\mathrm{av}}$) and minimal
          ($\mathcal{P}_{\mathrm{min}}$) output-state purities for different
          phases ($\phi$) \emph{with} feed forward ($p_\mathrm{succ}=50\,\%$).}
 \label{tab-withFF}
\end{table}

To evaluate how the feed forward affects the performance of the gate we have also calculated process fidelities, output state fidelities and output state purities for the cases when the feed forward was not active. It means we have selected only the situations when detector D$_{p0}$ (corresponding to $|+\rangle_P$) clicked and no corrective action was needed (like in Ref.\ \cite{mic08}). These values are displayed in Table~\ref{tab-withoutFF}. One can see that there is no substantial difference between the case \emph{with} the feed forward (success probability 50\,\%) and the case \emph{without} the feed forward (success probability 25\,\%).

\begin{table}
 \begin{ruledtabular}
  \begin{tabular}{cccccc}
  $\phi$ & $F_{\chi}$ & $F_{\mathrm{av}}$ & $F_{\mathrm{min}}$ &
  $\mathcal{P}_{\mathrm{av}}$ & $\mathcal{P}_{\mathrm{min}}$
  \\ \hline
  0         & 0.977 & 0.985 & 0.973 & 0.975 & 0.953 \\
  $\pi/6$   & 0.975 & 0.985 & 0.972 & 0.973 & 0.949 \\
  $\pi/3$   & 0.988 & 0.989 & 0.971 & 0.980 & 0.946 \\
  $\pi/2$   & 0.979 & 0.986 & 0.976 & 0.976 & 0.957 \\
  $2\pi/3$  & 0.981 & 0.989 & 0.966 & 0.982 & 0.935 \\
  $5\pi/6$  & 0.974 & 0.984 & 0.961 & 0.976 & 0.947 \\
  $\pi$     & 0.979 & 0.986 & 0.977 & 0.978 & 0.960 \\
  \end{tabular}
 \end{ruledtabular}
 \caption{Process fidelities ($F_{\chi}$), average ($F_{\mathrm{av}}$)
          and minimal ($F_{\mathrm{min}}$) output-state fidelities,
          average ($\mathcal{P}_{\mathrm{av}}$) and minimal
          ($\mathcal{P}_{\mathrm{min}}$) output-state purities for different
          phases ($\phi$) \emph{without} feed forward ($p_\mathrm{succ}=25\,\%$).}
 \label{tab-withoutFF}
\end{table}

~

\section{Conclusions}

We have implemented a reliable and relatively simple electronic feed-forward system which is fast and which does not require high voltage. We employed this technique to double the success probability of a programmable linear-optical quantum phase gate. We showed that the application of the feed forward does not affect substantially either the process fidelity or the output-state fidelities. Beside the improvement of efficiency of linear-optical quantum gates, this feed forward technique can be used for other tasks, such as quantum teleportation experiments or minimal disturbance measurement.

\bigskip

\begin{acknowledgments}
The authors thank Lucie \v{C}elechovsk\'{a} for her advice and for her help
in the preparatory phase of the experiment.
This work was supported by the Czech Science Foundation (202/09/0747),
Palacky University (PrF-2011-015), and the Czech Ministry of Education
(MSM6198959213, LC06007).
\end{acknowledgments}

\end{document}